\documentstyle[12pt,aasms4,rotating]{article}
\def\dm{{$\delta m^2  $}}
\def\dmatm{{$\delta m^2_{\small{ATM}}  $}}
\def\dmsun{{$\delta m^2_{\odot}  $}}
\def\dmsub{{$\delta m^2_{\small{SUB}}  $}}
\def\numu{{$\nu_\mu  $}}
\def\nue{{$\nu_e  $}}
\def\nutau{{$\nu_\tau  $}}
\def\nus{{$\nu_s  $}}
\def\RX{{SNR RXJ1713.7-3946}}

\def\simlt{\lower.5ex\hbox{$\; \buildrel < \over \sim \;$}}
\def\simgt{\lower.5ex\hbox{$\; \buildrel > \over \sim \;$}}

\def\gcm3{{\rm\,g\,cm^{-3}}}
\def\ncm3{{\rm\,cm^{-3}}}

\def\>{$>$}
\def\<{$<$}

\lefthead{Crocker, Melia \& Volkas}                      
\righthead{Supernova Remnant Neutrinos}

\renewcommand{\refname}{}   

\begin{document}

\title{\bf Discovering Long Wavelength Neutrino
Oscillations in the Distorted Neutrino 
Spectrum of Galactic Supernova Remnants}    

\author{Roland M. Crocker$^*$, 
Fulvio Melia\altaffilmark{1}$^{\dag}$ 
and Raymond R. Volkas$^*$}
\affil{$^*$School of Physics,					
Research Centre for High Energy Physics, 
\\The University of Melbourne,
3010 Australia\\
r.crocker, r.volkas@physics.unimelb.edu.au}
\affil{$^{\dag}$Physics Department and Steward Observatory, 
\\The University of Arizona, Tucson, AZ 85721\\
melia@physics.arizona.edu}

\altaffiltext{1}{Sir Thomas Lyle Fellow and 
Miegunyah Fellow}      



\begin{abstract}
We investigate the muon neutrino event rate in km$^3$ neutrino telescopes
due to a number of galactic supernova remnants expected on the
basis of these objects' known $\gamma$-ray signals. We evaluate the
potential of these 
neutrino signals to exhibit evidence of the sub-dominant neutrino
oscillations expected in various neutrino mixing schemes including 
pseudo-Dirac scenarios and the Exact Parity Model. With ten years' data,
neutrino signals from Sgr A East should either discover or exclude neutrino
oscillations governed by a $\delta m^2$ parameter in the range $10^{-12}$
to $10^{-15}$ eV$^2$. Such a capability is not available to terrestrial
or solar system neutrino experiments.
\end{abstract}




%

\section{Motivation and Plan}

In this work we present 
a novel extension of the basic idea that astrophysical
neutrinos can probe tiny values of the difference in the
squared masses of relevant neutrino mass eigenstates, \dm. 
This, namely, is that
through the observation of a 
deviation away from pure power-law scaling -- in other words, a
{\it spectral distortion} -- in a particular, galactic
supernova remnant's observed muon neutrino spectrum, an experimentalist
can infer the existence of exactly such a tiny mass
splitting. Such mass splittings are generic to a number
of extensions of the Standard Model such as the various active 
$\leftrightarrow$ sterile 
pseudo-Dirac scenarios and the Exact Parity Model.
The crucial advance in this proposed method is that {\it one does
not need to observe the species of neutrino into which the neutrinos emitted
at the source are oscillating} in order to diagnose oscillations. This is
important because both astrophysical \nue's and \nutau's are expected to be 
considerably more difficult to detect than astrophysical \numu's, at least
at more moderate energies.

The plan of the paper is as follows: in \S 
2 we discuss recent developments,
both empirical and theoretical, concerning galactic supernova remnants (SNRs),
including recent considerations of the factors limiting the
maximum energies to which they may accelerate particles.
In \S 3 we discuss neutrino production at SNRs through pion decay
and the relationship
between an SNR's $\gamma$-ray signal and its expected neutrino flux at Earth.
In \S 4 we briefly 
review neutrino oscillations and the status
of the various experiments purporting to demonstrate such oscillations. We also
introduce here the idea of tiny \dm's and review
their theoretical motivations. 
\S 5 considers the effect such tiny \dm's might have on the phenomenology
of SNR neutrinos. In \S 6 we briefly review the extensive code 
we have written
to model neutrino telescope detection of SNR neutrinos. Finally, we set out
the results of this code in \S 7 which goes on to
demonstrates the feasibility 
of employing the spectral distortion method briefly described above (and set
out in further detail later) to search for very long wavelength neutrino
oscillations. 

\section{Particle Acceleration in SNR Shells}
Cosmic-ray ions and electrons up to (and possibly exceeding) 
energies of
$\sim 10^{15}$ eV (near the so-called `knee' in the
distribution observed at earth) are widely believed to
be produced by galactic SNRs.  
The dominant acceleration mechanism
in SNRs appears to be diffusive (or first-order
Fermi) acceleration at the remnants' forward shocks.
Some support for this supposition was provided by the
EGRET experiment aboard the {\it Compton Gamma-Ray
Observatory}, which detected a large number of (at first)
unidentified sources in the super-$50$ MeV band, 
both in and above the Galactic plane.  Six of
these EGRET sources turned out to have compelling
associations with relatively young SNRs (\cite{esp}),
with gamma-ray fluxes at earth of typically a few 
$\times \ 10^{-7}$ photons cm$^{-2}$ s$^{-1}$ above $100$ MeV. 

The inference to be drawn here is that the accelerated ion (mostly proton)
and electron distributions of the EGRET SNRs
spawn neutral and charged pion decay
and the
accompanying bremsstrahlung, inverse Compton, and synchrotron
emission, that account for their broad-band photon spectra
from radio to gamma-ray energies.
In particular, the high energy photon spectra of these objects 
inferred from the EGRET 
measurements show
deviations from pure power-law scaling around 100 MeV that 
have been argued to be evidence of $\pi^0$ decay 
(\cite{prothsnr}; \cite{markoff}).
These  $\pi^0$'s 
would be expected from collisions between shocked
$p$'s in the SNRs' expanding  
shells and ambient nucleons.

EGRET has also detected
similar $\gamma$-rays from the Galactic Center (GC; \cite{mayer})
that 
Melia and others have strongly argued are associated with $\pi^0$ decay
at the SNR-like
object Sgr A East (\cite{melia}; \cite{markoff}).
(See Crocker, Melia and Volkas 2000
for an
analysis of the significance of Sgr A East  for $\nu$ 
astronomy.) 

The problem with
this otherwise straightforward interpretation, however, is that
a simple extension of the EGRET SNRs' GeV spectra to higher energies 
produces a TeV flux in excess of the upper limits established
by the Whipple (\cite{buckley98}; \cite{rowell}) and
High-Energy Gamma-Ray Array (HEGRA; \cite{prosch}) 
atmospheric
\v{C}erenkov telescopes and the CYGNUS extensive air shower array
(\cite{allen}; except in the case of the GC 
which {\it has} been seen -- marginally -- by Whipple; \cite{buckley}). 
Thus, a crucial question arises as to whether
shell-type remnants can indeed supply the observed Galactic
cosmic-ray population, while at the same time explaining
the emission from the handful of  unidentified, EGRET sources
associated with SNRs.  It may well be that the physical
conditions conducive to producing large numbers of relativistic
particles also  work against energizing any one of these
to TeV, let alone PeV, energies.

In their detailed analysis of this phenomenon, Baring et al. (1999)
adopted the view that the maximum energy attainable by ions in diffusive
shock acceleration is determined by 
 two complementary conditions, viz: (1) that the
acceleration time (as a function of energy) ought not to exceed the
age of the remnant (for the free expansion or early Sedov phase), and
(2) that the diffusion length of the highest energy particles ought not
to exceed some fraction of the shock radius. They concluded that,
for SNRs in a homogeneous environment, the circumstances that favor
intense gamma-ray production in the EGRET and sub-TeV bands (namely
an ISM density $> 1$ cm$^{-3}$) limit acceleration of particles 
to energies well below the cosmic-ray knee.  Thus, dense remnant
environments, as might be expected for the low Galactic-latitude
sources in the Whipple and HEGRA surveys, produce luminous emission
in the GeV range with a {\it simultaneous absence} of TeV $\gamma$-rays because
the resultant spectrum cuts-off (or significantly steepens) before this 
energy. 

At the same time, the detection of TeV emission from sources out
of the Galactic plane (and hence presumably in regions of lower ISM density)
leaves little doubt that there exists somewhat of an anti-correlation
between  GeV gamma-ray luminosity and super-10 TeV cosmic-ray
production in individual sources.  Neither the remnant of
SN 1006 (Tanimori et al.
1998), nor SNR RX J1713.7-3946 (Muraishi et al. 2000), were detected
by EGRET, yet their gamma ray signals were observed by CANGAROO.
(We present the
high energy $\gamma$-ray data for $\gamma$-Cygni and SN 1006
in Figure (1) for comparison.) 
Analysis of the  TeV gamma-ray emission from the northeast rim of 
SN 1006 indicates either super-$50$
TeV electrons inverse Compton scattering 2.7 K cosmic
microwave background photons or the decay
of $\pi^0$ mesons -- particularly if the remnant is close by
(\cite{aharonian}) -- or both. 
Note that the putative, high-energy
electrons might be either directly shock accelerated
or decay products resulting from collisions of shock-accelerated protons
and nuclei. Given that the $\gamma$-ray 
emission is from the 
rim, they may not, however, be 
accelerated by electro-magnetic processes associated 
with the remnant's neutron star.
In any case  -- {\it whether the gamma-rays are
hadronic or leptonic in origin} -- SN 1006 (and \RX \ ) present
evidence for shock 
acceleration of protons to high energies (at least $\sim 100$ TeV)
given that more significant
loss processes act on shock-accelerated 
electrons than protons in SNR environments. 

It appears, therefore, that should the cosmic-ray distribution be
produced by SNR shells, the most likely sources for the $\sim$ PeV
particles, like SN 1006, lie in regions of relatively low ISM density.
For SN 1006 
the TeV luminosity determined from the CANGAROO observations is
of order $10^{34}$ erg.s$^{-1}$, the exact figure depending on distance.  
Such energies are highly desirable for
the neutrino physics we wish to probe in this analysis. We
go on to calculate the level of neutrino emission from 
relevant SNR sources, which we expect will be predominantly out
of the Galactic plane.  



\begin{figure}[ht]\label{fig:difflux}
 {\begin{turn}{270}
\epsscale{0.8}  
\centerline{\plotone{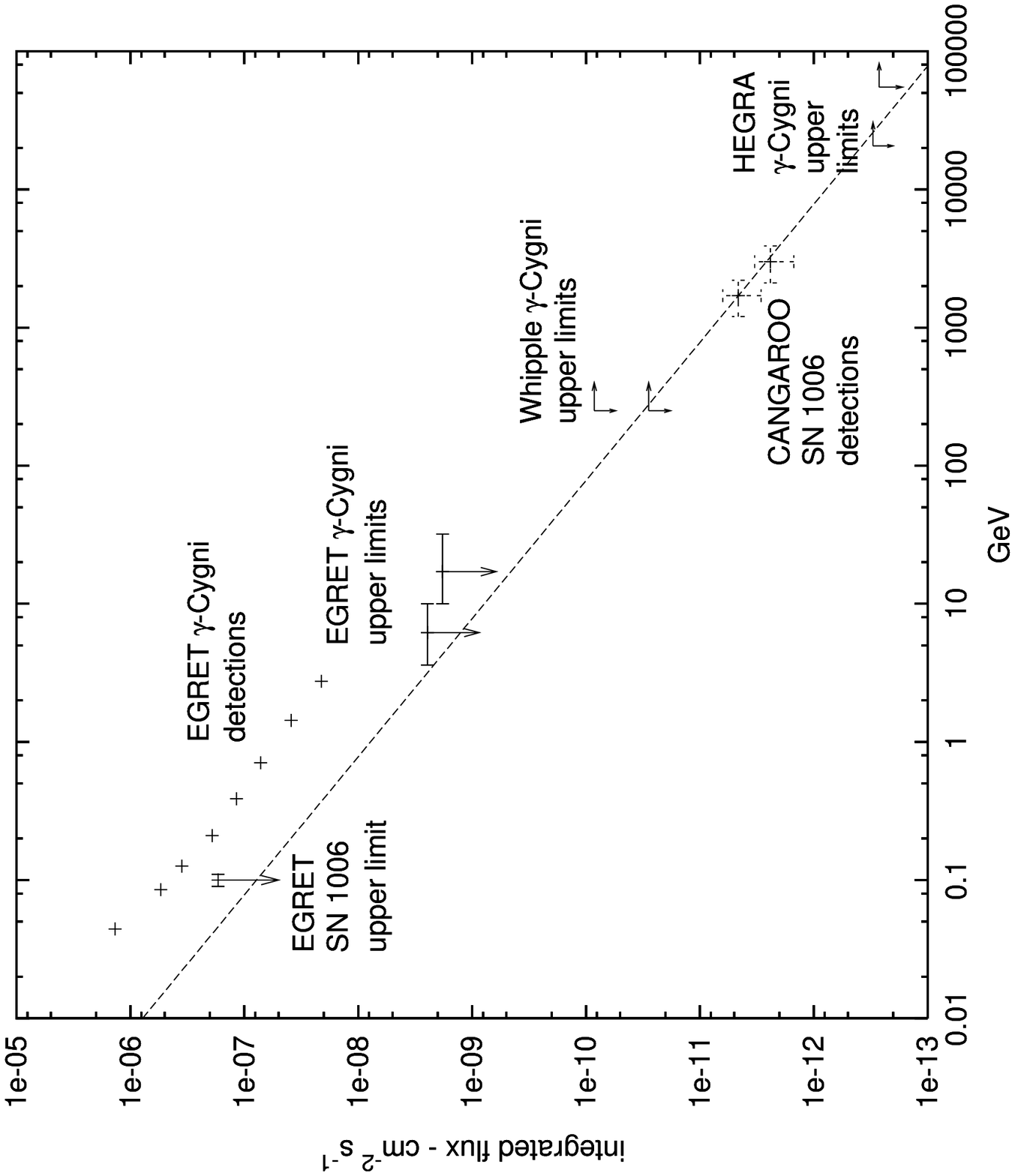}}
 \end{turn}}
\caption{$\gamma$-ray differential fluxes for $\gamma$-Cygni and SN 1006.
The EGRET data for $\gamma$-Cygni are from Esposito et al. (1996), 
the Whipple upper 
limits from Buckley et al (1998) and those due to HEGRA from 
Prosch et al. (1995). 
The SN 1006 data are from Tanimori et al. (1998) for the
CANGAROO collaboration and the EGRET upper limit is quoted by 
Aharonian and Atoyan (1999). The curve passing through the SN 1006 CANGAROO
data points is the expected integral flux from this object assuming
a spectral index of 2.0. This curve passes well below the
EGRET upper limit.}
\end{figure}

\section{Neutrino Production at Supernova Remnants}

Any proton-nucleon collision process at energies sufficient
to generate pions leads
to both $\nu_e$ and $\nu_\mu$
production from $\pi^\pm$ and $\mu^\pm$ decay; $\pi^\pm \rightarrow 
\mu^\pm \ \nu_{\mu}$ and 
$\mu^\pm \rightarrow e^\pm \ \nu_e \ \nu_\mu$. 
Note that we take $\nu$ to mean $\nu$ or $\overline{\nu}$
here, as we shall often do in the remainder of this paper,
because neutrino telescopes will not be able to distinguish
between $\nu$
and $\overline \nu$ at the energies attained by SNR neutrinos.
The ratio of neutrino flavors near the point of generation in an 
SNR will be very close to \nue \ : \numu \ : \nutau \ = 1 : 2 : 0,
as simple channel counting in the above
reactions would indicate. Certainly we 
expect no significant \nutau \ component at the SNRs (from charmed
hadron decay)
given the energies attained by the shocked protons (almost certainly no more
than 10 PeV). 

Given that these pion production and decay processes are well-understood
physics, the high-energy 
$\gamma$-ray signals from an SNR can, in fact, provide an empirical 
handle on its expected  total neutrino 
emission, $Q_{\nu}(E_{\nu})$.
In more detail,  $Q_{\nu}(E_{\nu})$
can be expressed as a function of the $\gamma$-ray emission at the SNR,
 $Q_{\gamma}(E_{\gamma}^0)$,
the index of the SNR's power-law proton spectrum above some energy 
($\alpha \simeq 2$
such as would result from shock acceleration), 
and $r \equiv ({m_\mu}/{m_\pi})^2$ (\cite{blasi}).
Note here that the decay neutrino spectrum will have the same spectral index,
$\alpha$, well above the $\Delta$ resonance domain, i.e., for energies 
$\gg 2 - 3$ GeV. 
We effect the relative normalization between the
$\gamma$-ray and neutrino distributions at $10$ GeV where
the pions take on a power-law form. 
From this we can derive a $\nu_{\mu}$
flux at earth and  
consequent event rate in a km$^3$ telescope (\cite{blasi}).

As a preliminary exercise then, one may determine whether 
the muon neutrino signal from 
a particular SNR is above atmospheric neutrino background at an energy
less than the maximum energy of the SNR $\nu$'s.
This is about $1/{12}$ the maximum energy to which the SNR might shock
$p$'s ($\sim$ 10 PeV)
given that $\langle x_F \rangle$, the average momentum carried by
the secondary pions relative to the parent proton is $\simeq 0.2$ 
(\cite{madph}; though note that
we shall find, in fact, given the expected power
law distributions of neutrino energies, that low statistics prevent 
observation of neutrino signals up to the maximum cut-off anyway). 
To perform this preliminary calculation, we 
assume a neutrino
spectrum of spectral index $\alpha \ = \ 2$ -- a value
well-justified theoretically for the case of shock acceleration
in SNR environments\footnote{Furthermore,
such a value is concordant with that expected for
the parent spectrum of galactic cosmic-rays as determined by
standard `leaky box' models (\cite{aharonian}).} 
and, in many cases able to be
inferred directly from an SNR's photon spectrum -- normalized to its
measured or inferred 10 GeV $\gamma$-ray signal. We also assume a
reasonable value of $\sim 1^\circ$
for a neutrino telescope's angular resolution.
Granted these inputs, we find that 
a window on the neutrino signal from these
objects is opened and we can consider what particle and astro- physics we
might extract from this ultra-high energy and ultra-long baseline neutrino
signal.

The
calculations we have performed show that 
SNRs typically emerge above the atmospheric
neutrino background at around a TeV, normalizing to their 
10 GeV $\gamma$-ray fluxes and assuming a 1 km$^3$ detector. 
Given that this is at or above the
energy regime 
at which the six EGRET SNRs (but not Sgr A East) can be expected to 
cut-out, we cannot safely predict
significant neutrino signal from these objects
in a km$^3$ detector.
On the other hand, the two CANGAROO sources can be expected to generate 
around 5 muon-like events per year in such a detector and Sgr A East 50
muon-like events, where
we have factored in a flux attenuation of one half
due to the averaged $\nu_{\mu} \rightarrow \nu_{\tau}$  
oscillations
we expect on the basis of the SuperKamiokande atmospheric neutrino
data (\cite{fukuda1}) -- see later. We now go on to consider the issue of
neutrino oscillations in somewhat greater detail.

\section{Neutrino Oscillations Between SNR and Earth}

\subsection{Distance Considerations}

The distance between the neutrino source and detector is about
8 kpc $\simeq 2.5 \times 10^{22}$ cm in the case of the Sgr A East. 
Distance 
determinations to
SN 1006 lie between around $0.7$
kpc (\cite{willingale})  to 2 kpc 
(\cite{winkler}). For
SNR RXJ1713.7-3946 they likewise range from  6 kpc 
(\cite{slane}) to 1
kpc (\cite{koyama}).

We take the GC neutrino 
source -- Sgr A East -- to have a linear dimension of around
$10\ pc \simeq 3 \times 10^{17}\ m$. Given the range of 
distance determinations, the linear dimensions of SN 1006 and 
SNR RXJ1713.7-3946 range between 6 and 27 pc and 7 and 41 pc respectively.
These distances are relevant because we need to know how neutrino
oscillation lengths compare with the size of the emitting object to
determine whether the neutrino source is flavor coherent.
If the former are small compared to the latter, then,
because neutrinos are emitted from points distributed across the source,
the oscillations will be averaged out. Alternatively, if the
latter are large compared to the former, then no averaging due
to the finite size of the source will be needed 
and the source is essentially
flavor coherent for neutrinos of a given energy.

Note that two
types of averaging generally need to be done: over distance, and
over energy. We need only consider distance 
averaging due to the finite size of the $\nu$ source. 
For SNRs the source distance 
scales involved are at least
six orders of magnitude larger than those for the detector (1 A.U. $\simeq
5 \times 10^{-6}$ pc). {\it Detector}-dependent
distance averaging, then, will not impact on calculations concerning 
neutrinos from SNRs. 
We do not address the issue of energy averaging due to
the finite energy resolution of the detector in great detail in this paper
(though we shall only ever consider the binning of event data into 
the coarse-grained scale of a decade of energy -- see later).

\subsection{Introduction to Neutrino Oscillations}

For the moment we consider 2-flavor oscillation modes $\nu_\alpha 
\leftrightarrow \nu_\beta$ for illustrative purposes.
Suppose a beam of flavor $\alpha$ is produced at $x = 0$.
Then at a point $x$ distant from the source the oscillation
probability is
\begin{equation}
P(\alpha \to \beta) = \sin^2 2\theta \sin^2 \left(
\pi \frac{x}{L_{osc}} \right). \label{osceqn}
\end{equation}

\noindent The parameter $\theta$
is the `mixing angle' which determines
the amplitude of the oscillations. 
The value $\theta = {\pi}/{4}$, which leads to the largest possible
amplitude, is termed `maximal mixing'.
The parameter $L_{osc}$
is the `oscillation length' -- the length scale over
which oscillation between two mass eigenstates occur -- and is given by
$L_{osc} = \frac{4 \pi E}{\delta m^2}\hbar c$. 
This works out to be
\begin{equation}
L_{osc}  \simeq 0.8 \frac{E/PeV}{\delta m^2/10^{-10}\ eV^2} \ kpc.
\label{oscform}
\end{equation}
Note that the oscillation length increases linearly
with energy. 
The parameter
$\delta m^2 \equiv |m_1^2 - m_2^2|$ is the
squared-mass difference between the two mass 
eigenstate neutrinos. 

Totally averaged oscillations see the $x$-dependent $\sin^2$
factor in Equation (\ref{osceqn}) set equal to ${1}/{2}$, leading to
\begin{equation}
\langle P(\alpha \to \beta) \rangle = \frac{1}{2}
\sin^2 2\theta.
\end{equation}
This, to reiterate, can be due to either distance or
energy spread or both.

We now briefly review the various possible solutions to
the atmospheric and solar neutrino problems
before considering the issue of sub-dominant neutrino oscillations.

\subsection{Atmospheric Neutrinos}
SuperKamiokande detects a $50\%$ deficit of $\mu$-like atmospheric neutrinos
coming up through the earth (\cite{fukuda1}).
They see no deficit of either
upward- or downward-going 
$e$-like neutrinos. The lower energy, downward-going $\mu$-like
events are deficient, whereas their high-energy counterparts are not.
These data can be explained by close-to-maximal $\nu_{\mu} \to \nu_x$
oscillations with $x \neq e$ and $x = \tau$ or $x = s$ (sterile), or a
combination thereof.
These two alternatives both require parameters in {\it approximately}
(see, e.g., \cite{foot}; \cite{fornengo})
the same range (\cite{atmos}): 

\begin{equation}
\delta m^2_{\small{ATM}} =
10^{-2}\ eV^2 \to 10^{-3} \quad {\rm and}\quad 
\sin^2 2\theta_{\mu x} = 1.  \label{mutau}
\end{equation}

\noindent
SuperKamiokande currently favors oscillations to
$\nu_\tau$ over oscillations to a sterile neutrino at the 2$\sigma$ level
(\cite{fukuda2000}), though some have cast doubt over this result
(\cite{foot2};\cite{kobayashi}). Furthermore, fits to oscillation to 
a combination of \nutau \ and \nus \ 
fit the data equally well as oscillation to \nutau \ alone
(\cite{fogli}).
In this work, however, for definiteness
we will take it that the resolution of the
atmospheric neutrino anomaly lies in maximal 
$\nu_\mu \to \nu_{\tau}$ oscillations over the
relevant length scale, $L_{ATM}$.

Substituting \dmatm \ into Equation (\ref{oscform}), we see that the 
$\nu_\mu \to \nu_x$ oscillation length is 
orders of magnitude less
than the size of a typical SNR shell
for the entire neutrino spectrum (i.e., up to an energy of $\sim$ 1 PeV).
This means that the oscillations will be distance averaged,
and hence at earth we expect
a $50/50$ mixture of $\nu_\mu$
and $\nu_{\tau}$ from a SNR.

\subsection{Solar Neutrinos}
The solar neutrino problem can be solved by $\nu_e \to \nu_y$
oscillations, where $y = \mu, \tau, s$, or a combination
thereof, are all allowed, with
one important proviso: if the Los Alamos LSND experiment is
correct, then $\nu_e \to \nu_\mu$
oscillations, with parameters that cannot solve the solar
neutrino problem, have already been detected (\cite{LSND}).
So, if the still-controversial LSND result is correct, 
then $y = \mu$
is ruled out. The MiniBOONE and BOONE
experiments at Fermilab should eventually settle 
this issue (\cite{bazarko}). In this work we disregard the LSND result,
though in the generic six neutrino mass eigenstate model discussed below it 
could be easily accommodated.

The precise oscillation parameter space required to account for
the solar data depends on which of the solar neutrino
experiments are held to be correct. The two parameter ranges defined
 below, however, are broadly consistent with all solar data; 

\begin{enumerate}
\item $\nu_e \to \nu_y$ with a {\it small} mixing
angle (SMA) $\theta_{ey}$ is possible through the MSW
effect. If this pertains, then the oscillation
amplitude will be far too small to affect SNR
neutrinos, but the most recent data from SuperKamiokande disfavor
this solution (\cite{SuperK2001}). We therefore disregard this possibility.
\item $\nu_e \to \nu_y$ with a very large mixing angle
(LMA) $\sin^2 2\theta_{ey} \simeq 1$ is an interesting
possibility for the range\footnote{Note, however, that some of this
range is excluded because of the non-observation of a `Day-Night Effect'
in the solar neutrino signal, i.e.,
the absence of the diurnal variation due to the MSW-induced `regeneration'
in the earth
 expected if \dmsun \ were in the
$\sim 10^{-6}$ eV$^2$ range and maximal mixing applies (\cite{guth};
\cite{crocker1}; \cite{smoking}.} 
\begin{equation}
\delta m^2_{\odot} = 10^{-3} \rightarrow 10^{-10} \ eV^2.
\end{equation}
\end{enumerate}
The lower end of the solar 
\dm \ parameter space, $\delta m^2_{\odot} \lesssim 10^{-9}\ eV^2$ 
defines
`just-so' oscillations where the oscillation
length for solar neutrinos 
is of order 1 A.U. $ \ $For larger $\delta m^2_{\odot}$ \
values completely averaged oscillations,
with a flux suppression factor of $1/2 \sin^2 2\theta_{ey}$, result.
Maximal mixing explains almost all
of the data with averaged oscillations (excepting
 the Homestake result; \cite{cleveland}). 
Values of $\delta m^2_{\odot} > 10^{-3}\ eV^2$ are ruled out by
the non-observation of $\overline \nu_e$ disappearance
from reactors (CHOOZ; \cite{chooz} and Palo Verde experiments; \cite{boehm}).

If the solar problem is resolved with  $\delta m^2_{\odot} \lesssim 10^{-8} 
eV^2$ range, one finds from 
Equation \ref{oscform} that
$L_{\odot}$
becomes larger than an SNR with typical linear dimension $\sim 10$ pc.
within the
energy range spanned by SNR neutrinos.
This means that above some critical energy the
$\nu_e$ beam from such a source would be flavor-coherent (of course, 
conversely for larger \dmsun's, SNR \nue's would be distance-averaged
over their entire energy range),

In principle, such coherence
would evidence itself by an energy dependent spectral distortion; 
the $\nu_e$ flux 
at a particular energy ($E \to E+\Delta E$) 
would depend on the part of the neutrino
oscillation wave (for that particular energy) encountered by the 
earth at its distance from the particular SNR. Pragmatically, 
given the small statistics 
that will accrue from the SNR neutrino sources and the limited energy 
resolution expected to be achieved by any of 
the proposed neutrino telescopes, 
one expects no observational consequence of this flavor coherence. 
This is because the energy dependence of the flux suppression
washes out with the inevitably large size of the energy bins
particular neutrino events are accumulated into.
The beam, therefore, 
is indistinguishable from one in the distance-averaged 
oscillation regime.

Again with reference to Equation (\ref{oscform}),
note also that at the {\it extreme} lower end of the allowable
parameter space, $\delta m^2_{\odot} = 10^{-10}\ eV^2$,
for a close-by SNR at 1 kpc and at a neutrino energy of 1 PeV,
the $\nu_e \to \nu_y$ oscillation length, $L_{\odot}$,
reaches the order of the SNR-earth distance, $L_{SNR}$. 
The hypothetical SNR's $\nu_e$ flux at the earth would, then,
rise from being suppressed somewhat below a PeV to unsupressed
somewhat above a PeV if the $\nu_e$'s attained this energy. 
Unfortunately, we do not reasonably expect
any observational consequence from this potentially interesting phenomenon;
in the first place, we would be relying on an unlikely
coincidence of extreme energy and absolutely minimal allowable \dm,
and secondly, neutrino telescopes almost certainly will
not be able to detect \nue's of SNR energy.

This \nue-blindness basically 
stems from the fact that electrons resulting from
charged current 
$\nu_e$ events of PeV energy and below
are
arrested extremely quickly in the detector medium. 
Such \nue's are also well beneath
the Glashow resonance 
(\cite{glashow}; \cite{berezinsky}; \cite{gandhi1}) and also the energy
thresholds of `alternative' astrophysical neutrino detection techniques 
that may be suitable for ultra high-energy astrophysical \nue \ detection
like air shower arrays (\cite{capelle})
and radio detection of neutrino interactions
in Antarctic ice 
(\cite{gaisser}; \cite{alvarez1}; \cite{alvarez2}).
The short electron path length leaves the
experimentalist with essentially no directional information
to distinguish signal and background.
Furthermore, even if 
directional information could be somehow obtained,
the electromagnetic showers generated by braking electrons
and the accompanying hadronic showers from the charge current interactions
with nuclei in the detector medium
are essentially indistinguishable
from the purely hadronic showers resulting from neutral current interactions
of all $\nu$ flavor at these energies.
They are also difficult to distinguish from lower energy
($E_{\nu_\tau} < 100 \ TeV$)
\nutau \ charged current events (\cite{ANTARES}).
In such events 
the exiting $\tau$ does not have a large enough $\gamma$ factor 
for it to decay sufficiently far from the 
interaction vertex for the `double bang' (\cite{learned}) -- the hadronic
showers due to the original charge current interaction and the later
$\tau$ decay -- to be resolved (see \cite{crocker2} and
references therein for more detail
on the chances for SNR-energy \nue \ detection). 

Given that double-bang detection of \nutau \ also only becomes practicable
in the uppermost decade of the SNR neutrino energy spectrum
(\cite{ANTARES}),
in this work
we will only consider possibilities for SNR $\nu_\mu$  detection, 
though note that SNR \nue's remain of
some importance because of what they might oscillate {\it into}.


\subsection{Atmospheric and Solar Data Combined}

Note firstly that any neutrino
mixing scheme that seeks to accommodate
both the atmospheric and solar neutrino
anomalies, with their different
attendant oscillation length scales, and, therefore, different \dm's,
must possess {\it at least} three distinct neutrino mass eigenstates.
Furthermore, with just the minimal three neutrino mass eigenstates, 
if we demand maximal 
\numu $\rightarrow$ \nutau \ oscillations over the atmospheric
scale and maximal mixing also in the solar case,
we find that \nue \ is forced to be maximally mixed with both \numu \ and
\nutau \ over solar length scales. In other words, we have
bi-maximal mixing (\cite{bimax1}; \cite{bimax2};
\cite{bimax3}; \cite{bimax4}; \cite{bimax5}; \cite{bimax6}).

That close-to-maximal mixing is demanded by the 
atmospheric neutrino anomaly (\cite{fukuda1}; \cite{apollonio})
and also favored by the most recent solar neutrino 
data (\cite{SuperK2001}) raises hope for
extracting interesting particle physics from
astrophysical neutrino phenomenology.
This is because the poor statistics of the proposed
neutrino telescopes mean that only modes with large mixing angles,
$\theta$, can in practice be probed. 

\subsection{Finding Oscillations}
Generically,
neutrino oscillations may be evidenced in three ways, viz:
\begin{enumerate}
\item  By a difference between the neutrino 
flavor ratios at point of generation to those determined 
at point of detection, i.e., in the case of
SNRs, by a difference
between \newline \nue \ : \numu \ : \nutau \ = 1 : 2 : 0 at source and
\nue \ : \numu \ : \nutau \ = $1\times P(\nu_e \to \nu_e) \ + \ 2\times
P(\nu_\mu \to \nu_e) \ : \ 1\times P(\nu_e \to \nu_\mu) \ + \ 2\times
P(\nu_\mu \to \nu_\mu) \ : \ 1\times P(\nu_e \to \nu_\tau) \ + \ 2\times
P(\nu_\mu \to \nu_\tau)$ at detector. In practice, this could probably
only be probed by 
measurement of ${\Phi_{\nu_{\tau}}^{obs}}/{\Phi_{\nu_{\mu}}^{obs}}$
(given the difficulty with astrophysical \nue \ detection outlined already),
which in the absence of oscillations should be zero. 
\item  Via
an observed discrepancy between detected and expected flux,
given one has an accurate fix on the absolute flux
of a particular neutrino species (which, as already discussed, could not
be \nue \ in the SNR case)
expected on the basis of an
SNR's $\gamma$-ray signal: 
${\Phi_{\nu}^{obs}}/{\Phi_{\nu}^{theor}} \ \neq \ 1$, where
$\Phi_{\nu}^{obs}$ denotes the observed flux
of SNR \numu's or \nutau's \ and 
$\Phi_{\nu}^{theor}$ denotes the flux of either flavor expected
in the absence of oscillations.

Essentially, the observation of \nutau's from an SNR constitutes
evidence for oscillations under either oscillation diagnostic (1) or (2).

\item Through the observation of a {\it spectral anomaly}, i.e., a
distortion of a particular neutrino flavor's
energy distribution  
away from its expected 
shape. In the case of an SNR's $\nu$ signal, this would mean a deviation
away from pure power-law scaling of neutrino flux with energy (given that
it could be determined {\it a priori} that the region of the distribution
under investigation should be governed by a single spectral index). The great 
advantages of this third method over the former two are that: 

\noindent i) it only 
requires observation of a single neutrino species: \numu's 
for SNRs in practice. 
One is
{\it not} obliged to positively identify the $\nu$ flavor to which the
$\nu_\mu$'s might oscillate to uncover oscillation evidence nor is one
required to know precisely what the expected flux of \numu's is; 

\noindent ii) it can give a {\it range} for the \dm \ governing the
oscillations, not just a lower bound.
\end{enumerate}

\subsection{Sub-dominant Oscillations}

Now, we have already seen that 
for the entire allowable $\delta m^2_{\small{ATM}}$ and $\delta m^2_{\odot}$
regimes we pragmatically expect 
totally averaged oscillations in SNR signals, 
i.e., precisely half the
$\nu_e$'s and $\nu_{\mu}$'s generated in a SNR
should oscillate to something else on their
journey to the earth given bi-maximal mixing (though the actual number
of \nue's detected would not vary from the naive expectation because of 
oscillations {\it from} the maximally mixed \numu --\nutau \ sub-system to
\nue -- see, e.g., \cite{bento}; \cite{athar}). 
In other words, the three oscillation diagnostics outlined above,
 when applied to
SNR neutrino signals,
would only ever act as confirmatory to existing
terrestrial neutrino oscillation experiments, albeit over  
vastly different energy and distance regimes. In more detail,
with the three mass-eigenstate 
bi-maximal mixing outlined
above, we expect (i)\nue \ : \numu \ : \nutau \ = 1 : 1 : 1 at detector
with the (possibly) practical, phenomenological implication that 
${\Phi_{\nu_{\tau}}^{obs}}/{\Phi_{\nu_{\mu}}^{obs}} \simeq 1 $, (ii)
${\Phi_{\nu_{\mu}}^{obs}}/{\Phi_{\nu_{\mu}}^{theor}} \ = \ 1/2$ independent
of energy, and (iii)
{\it no} spectral distortion of the \numu \ signal (i.e., a plot of
neutrino flux versus energy on a log-log scale simply produces a line
of constant slope).

Note, however, from Equation (\ref{oscform}), that, if an extra \dm \
significantly smaller than $10^{-10}$  eV$^2$ 
were operating in astrophysical neutrino oscillations
-- thereby introducing an extra, longer oscillation length scale --
things become interesting. In particular,
an energy dependent {\it spectral distortion} might be 
observed in the  SNR's \numu \ signal - see later.
We label this  new \dm,
\dmsub,  where `SUB' denotes {\it sub-dominant} oscillations.


Of course, 
any scenario invoking this new oscillation scale
demands
one or more
new neutrino mass eigenstates. 
This results in the presence in the theory
of additional weak eigenstates which
are obliged to be sterile neutrinos. Of course,
these new weak eigenstates present us with
alternative possibilities for resolving one or other of the
existing neutrino anomalies, e.g., we might
solve the solar anomaly with \nue $\rightarrow$ 
\nus \ oscillations. 

Consider momentarily the range
of \dmsub \ potentially probed by SNR neutrino signal. To see the spectral
distortion mentioned above we require that it occurs,
for a particular SNR at some distance $L_{SNR}$,  at an energy
{\it below} the maximum attained by the SNR neutrinos
(in fact, somewhat below given
the expected tail-off of high energy events with a power-law spectrum)
and {\it above} the energy at which the SNR's signal becomes invisible
because of atmospheric neutrino background. As mentioned
previously, 
our calculations 
show that this energy is around a TeV for all nine SNRs so far discussed.
The \dmsub \ ranges
probed, with such a threshold energy, are approximately 

\begin{equation}
10^{-9} \rightarrow 10^{-13} {\rm eV}^2 \ {\rm for \ 1 \ kpc}
\quad {\rm and}\quad 
10^{-10} \rightarrow 10^{-14} {\rm eV}^2 \ {\rm for \ 10 \ kpc.}
\label{range}
\end{equation}

One may note immediately that the \dmsub \ ranges discussed, given
they {\it are} so tiny, are not ruled out by any existing
neutrino oscillation experiment (even with the largest
\dmsub, at an energy of 1 MeV, $L_{SUB}$ is still over a million
kilometers). Put another way, {\it if such \dmsub \ scales operate in nature,
we} can {\it only probe them with astrophysical neutrinos}. 
These values for \dmsub \ are tiny
numbers. We now go on to discuss how
we might motivate them.

\subsubsection{6-Neutrino Maximal Mixing Schemes}

As in the case of solar and atmospheric scale oscillations, for there
to be an observational consequence of  sub-dominant oscillations for
astrophysical neutrinos, we require that the mixing 
behavior described immediately above 
be close-to-maximal. We therefore seek neutrino mixing scenarios
that not only naturally incorporate tiny mass splittings, but also result
in maximal or close-to-maximal mixing.
There are, indeed, two examples of such that we know of: 
the Exact Parity Model (\cite{mirror1}; \cite{mirror2};
\cite{mirror3}; \cite{mirror4}) and
the generic,
active $\leftrightarrow$ sterile,
pseudo-Dirac scenario (\cite{wolfenstein};  \cite{bilenky}; \cite{bilenky2}
\cite{kobayashi2};
\cite{giunti};
\cite{bowes};
\cite{geiser};
\cite{kobayashi}).
In both these scenarios pairwise  maximal and close-to-maximal
(respectively) mixing between every active neutrino flavor and its sterile
partner can naturally explain the solar and atmospheric anomalies
with \nue $\rightarrow$ \nue$'$ and \numu $\rightarrow$ \numu$'$. 
Alternatively,
both can also accommodate close-to-maximal
intergenerational mixing
so that, for instance, the atmospheric anomaly be resolved by
\numu $\rightarrow$ \nutau \ oscillations (\cite{yoon}; \cite{kobayashi}), 
as we assume in 
this work, whereas the
solar neutrino anomaly continues to be explained by \nue $\rightarrow$ \nue$'$,
i.e., active to sterile,
oscillations. In this latter situation, demanding that 
the scale of the mass-splitting
between \numu \ and \numu$'$ does not interfere
with atmospheric neutrino experiment results only constrains  \dmsub \ to 
be somewhat less than $10^{-3} \ eV^2$. It is natural, however, to assume that
the mass splitting between \numu \ and \numu$'$, \dmsub,  be of the order of
that between \nue \ and \nue$'$, \dmsun.

\subsubsection{4-Neutrino Maximal Mixing Schemes}

We are not aware of any particularly strong theoretical motivations for
maximal mixing in a four neutrino system with very small mass-splitting, but
we consider this generic case for completeness. 
Introducing
a fourth, light neutrino mass eigenstate with a mass
very close to one of the existing states will result in sub-dominant
oscillations of both \numu \ and \nue \ to a new \nus \ 
over long length scales.
Again if this mass difference is in the \dmsub \ range described above and 
the mixing is maximal, there will be phenomenological consequences for 
astrophysical neutrino signals from SNRs.
 
With three mass splittings, \dmatm, \dmsun, \ and \dmsub \ and four
mass eigenstates, there are six
possible arrangements of the latter. These can be broken down into
two `double-doublets' (arrangements of two pairs of mass eigenstates
defining \dmsun \ and \dmsub \ split by \dmatm) 
and  four `mixed' arrangements.
 
\section{Observational Consequences of Oscillation Scenarios}

\subsection{ In Theory}

The phenomenological consequences of 
the maximal, sub-dominant 
four and six mass eigenstate
scenarios mentioned
can be gauged by noting the energy dependence that they introduce to
$F_{\nu_\mu}$, the fraction of the total initial neutrino flux from an SNR
that arrives at earth with flavor \numu:
\begin{equation}
F_{\nu_\mu} \ = \ 1/3 \times P(\nu_e \to \nu_\mu) \ + \ 2/3 \times
P(\nu_\mu \to \nu_\mu) \ \equiv \ \rho - \sigma \sin^2 \Delta(E)_{SUB},
\label{oscparam}
\end{equation}
\noindent where we define 
$\Delta_{SCALE} \equiv \frac{\delta m^2_{\small{SCALE}} \ L_{SNR}}
{4E \ \hbar c}$ and $\tiny{SCALE} \in \{\tiny{ATM}, \odot, \tiny{SUB} \}$. 
$\rho$ is $1/3$ for all the sub-dominant oscillation
scenarios under investigation whereas
$\sigma$ is $1/3$ in the case of the six mass eigenstate
scenario, $1/6$ for the two, four mass eigenstate
 double-doublet scenarios, and $1/12$ in the
case of the four,  four mass eigenstate mixed scenarios.

The above expression for $F_{\nu_\mu}$ is, in principle, dependent
on all three oscillation scales -- $\Delta_{SUB}, \Delta_{\odot}$, and 
$\Delta_{ATM}$. In practice, no dependence of $F_{\nu_\mu}$ on 
$\Delta_{ATM}$ is evident because, over the entire energy range of
any SNR, atmospheric oscillations will be averaged as already discussed.
More interestingly, any dependence on $\Delta_{\odot}$
actually cancels out between $1/3 \times P(\nu_e \to \nu_\mu)$ and 
$2/3 \times P(\nu_\mu \to \nu_\mu)$ essentially because \nue \ and \numu \ are
maximally mixed over the solar and atmospheric scales respectively.

The energy dependence in $\rho - \sigma \sin^2 \Delta(E)_{SUB}$ 
will show up
in the latter two oscillation diagnostics provided sufficient
statistics can be accrued and \dmsub \ falls within the range defined by
(\ref{range}): 
 over
the energy range of an SNR's detected $\nu$ \ spectrum, 
${\Phi_{\nu_{\mu}}^{obs}}/{\Phi_{\nu_{\mu}}^{theor}}$
will go from
some constant fraction ($< 1/2$) well below $E_{crit}$, to exhibiting 
oscillatory behavior around $E \ = \ E_{crit}$, to a constant value of
1/2 well above $E_{crit}$. Also, in these three regimes, (over
increasing energy) a plot of the SNR's differential
$\nu$ flux 
versus energy on a log-log scale  produces a line which at first has
some constant slope, $\alpha$, goes through some oscillatory
regime around  $E \ = \ E_{crit}$
and then resumes along the initial constant slope, $\alpha$.
By $E_{crit}$ we denote the energy at which
$L_{SUB}(E_{crit}) \ = \ L_{SNR}$, where  
$L_{SUB}(E) = \frac{4 \pi E}{\delta m^2_{\small{SUB}}}\hbar c$ and we have
employed Equation
(\ref{osceqn}) setting $\theta = \frac{\pi}{4}, \ x = L_{SNR} \ \& \  
L_{osc} = L_{SUB}$.

Note that because $\sigma$ is constant within each categorization 
(six mass eigenstate and four eigenstate, either 
`double-doublet' or `mixed'),
different examples within each of these categorizations produce the 
same phenomenological consequences in terms of the latter two oscillation 
diagnostics.

The first oscillation diagnostic does not 
identify sub-dominant oscillations to a \nus, however, because with
\numu \ and \nutau \ already mixed over the atmospheric scale, we
simply expect
${\Phi_{\nu_{\tau}}^{obs}}/{\Phi_{\nu_{\mu}}^{obs}} \ = \ 1/2$,
independent of energy.

\subsection{In Practice}

Of course, neutrino telescopes do/will not measure neutrino fluxes directly.
Rather they seek to pick out -- from the \v{C}erenkov radiation they emit --
the limited number of charged leptons that cross
their instrumented volumes of ice or water 
that are due to charged current interactions of
astrophysical neutrino near or in these detector volumes. 
This signal must be extracted from the massive background of down-going,
atmospheric muons that also cross the detector volume.

In order to show that the techniques we have described above 
can be  useful diagnostics
for sub-dominant oscillations in SNR $\nu$ signals we must determine whether
the actual event rates in a km$^3$ detector due to promising  SNR $\nu$ 
sources are large enough that plots of neutrino flux versus
energy for these sources have small enough statistical errors that
deviations from pure power-law scaling (i.e.,
deviations from linearity on a log-log scale) due to sub-dominant oscillations 
might be positively identified against unavoidable statistical fluctuations.
The derivation of fluxes from event rates also requires that
a detector's energy-dependent response function be sufficiently well
characterized and we assume this is the case.

There are plans afoot for the construction of km$^3$-scale
neutrino telescopes in both the Antarctic ice (AMANDA and its planned
extension
IceCube) and the deep
Mediterranean sea (the ANTARES and NESTOR projects).
For recent reviews of the status of neutrino telescopy 
and descriptions of the various neutrino telescope proposals
the reader is directed to Spiering (2001) for the IceCube project,
The ANTARES Collaboration (1999), and Botai et al. (2000) for NESTOR. 

\section{Simple Modeling of Detector Operation}

The probability that a high
energy muon-type 
neutrino is detected in a $km^3$-scale neutrino telescope depends on
two factors, viz; approximately inversely on the interaction
length of the neutrino ($\lambda_{int}$) at 
that energy (which, in turn, depends on the charged current cross-section) 
and approximately directly on the 
radiation length of the muon ($R_\mu$) 
 produced in the interaction (\cite{halzen}). 
(We assume here that the linear dimension of the detector is small
on the scale of $R_\mu$.)  
We can make a rough estimate 
of the effect of these factors by writing down a detection probability 
multiplier which goes as some power of the energy:
$$
P_{\nu \rightarrow \mu} \simeq \frac{R_\mu}{\lambda_{int}} \simeq A {E_\nu}^n.
$$
In our model we employ the values for $n$ and  $A$ given by Halzen (1998).

We have written a FORTRAN code which, from a particular SNR's 10 GeV 
$\gamma$-ray
flux, extracts event rates in a neutrino telescope of $1 \ km^3$ volume,
assuming that the $\gamma$-ray emission is hadronic in origin. The code
takes into account earth shadowing effects relevant to a particular SNR
given its declination and the detector's latitude.
These we estimate on the basis of the work of Naumov and 
Perrone (1999). The event
rates are given per decade energy bin from $10^2 - 10^3$ GeV to 
$10^5 - 10^6$ GeV.
The code determines these event rates given various values for \dmsub \ and 
for the minimum and maximum  determined distances to each SNR.
The code does {\it not} assume that the neutrino signal has to be 
up-coming for it
to be detected. Rather, for nadir angle bins from $0 \to 10^\circ$ 
to $170 \to 180^\circ$
it compares the signal (more precisely, the differential neutrino flux 
due to the SNR) to the atmospheric background over an assumed detector
resolution.
This background has two components: atmospheric \numu's \ and, for
nadir angles greater than $90^\circ$, atmospheric $\mu$'s.
We employ the zenith angle-dependent
parameterization of the sea-level \numu \ and $\mu$ flux
given by Lipari (1993) and also employ the results of  Lipari and Stanev
(1991) to
account for the attenuation of muons with their propagation through the Earth.
In the muon-attenuation sub-routine we have made
the approximation that each detector is located at a single,
well-defined depth below the Earth's surface (1.6 km in the case of a
South Pole and 2.4 km in the case of a Mediterranean detector).
The code starts to record
events in a particular, 10$^\circ$
angle bin (of which, of course, there are 18 within
each energy bin) only when the signal rises above the background. Because the
atmospheric \numu \ and $\mu$ backgrounds go with spectral index $\alpha = 3.7$
in the energy ranges of concern, whereas the sources go with index 
$\alpha \simeq 2$, the background quickly drops away from the signal once it
has been surpassed. 

We estimate the detector angular resolution by employing
the parameterization suggested by the ANTARES Collaboration (1999):
\begin{equation}
\Theta \ = \ \frac{0.7^\circ}{(E_\nu / TeV)^{0.6}} \ + \ 0.1 ^\circ.
\label{res}
\end{equation}
This function estimates the importance of the 
three factors that limit the determination of the primary neutrino's
direction of travel. These are the uncertainty in the angle between the
incoming $\nu_{\mu}$ and the resulting $\mu$, the deviation 
of the $\mu$ away from its original direction of travel due to 
multiple scattering and, lastly, the detector's intrinsic angular resolution
as determined by uncertainties in its exact geometry, etc.
The ANTARES collaboration (1999) has
determined from Monte Carlo 
simulations that below $10$ TeV total angular resolution
is limited by the unavoidable angular distribution of the neutrino 
interactions whereas above $100$ TeV it is limited by detector effects. 

The AMANDA project (which will hopefully evolve into IceCube) 
has to contend with the short scattering length of the \v{C}erenkov
light in ice, $24$ m , as compared to sea water at greater than $200$ m. 
Despite this, IceCube will achieve an angular resolution less than one 
degree and perhaps as low as 
$0.4^{\circ}$ (\cite{halzencomm}) and we, perhaps optimistically,
adopt the same parameterization of detector resolution -- 
Equation (\ref{res}) --
independent of detector medium (i.e., ice or water).

The binning of event rate data into decade energy bins is
forced upon us 
by two factors: the expected limits to the energy resolution of the 
proposed detectors and low statistics. Again, 
accurate determination of the energy possessed by a muon neutrino (which
produces a muon observed by a detector) is limited by three factors: 
uncertainty in the fraction of the neutrino's total energy 
imparted to the muon, ignorance of the energy loss by the muon outside the
instrumented volume and, finally, the intrinsic 
energy resolution of the detector apparatus
itself (\cite{ANTARES}). 
Given these three factors, the ANTARES collaboration (1999) has judged
on the basis of MC simulations of their detector array that they 
can gauge a muon neutrino's energy to within a factor of three for 
$E_{\nu} > 1$ TeV, so binning 
simulated data
into decades of energy is
reasonable.

\section{Results}

Because of their locations in southern skies, the three SNR-like objects
that we have determined should produce detectable neutrino fluxes at the 
earth -- the Sgr A East, 
SN 1006 and \RX \ -- must exceed the atmospheric {\it muon}
background before they can be visible to IceCube. This means that,
whereas all three emerge above background at or below $\sim$ TeV  for a 
detector at Mediterranean latitudes (where they are
shielded from muons for at least part of the day), 
at the South Pole their signals only emerge above $\sim$ 10 TeV. 
The ten-year, no-oscillation event rates in a km$^3$ detector, per decade
energy bin we determine for these three
objects are presented in Table (1).
\begin{table}[h]
\begin{tabular}{|c|c|cccc|} \hline
detector & object & $10^2 \to 10^3$ & $10^3 \to 10^4$ & 
$10^4 \to 10^5$ & $10^5 \to 10^6$ \\ 
location & & & & & (in GeV) \\
\hline
Mediterranean & & & & & \\
 	& Sgr A East	& 108 	& 491 	& 277 	& 118	\\
	& SN1006& 0	& 18	& 16	& 8 	\\
  	& SNR RX...	& 0	& 23	& 20	& 10	\\ 
\hline
South Pole & & & & & \\
 	& Sgr A East	& 0 	& 0 	& 156	& 152	\\
	& SN1006& 0	& 0	& 1	& 11 	\\
  	& SNR RX...& 0	& 0	& 3	& 14    \\
\hline

\end{tabular}
\caption{Event rates in decade energy bins due to the three sources
under consideration assuming no oscillations. 
}
\end{table}

\vspace{0.5cm}

\noindent These figures should be 
halved to incorporate the effect of averaged
\numu $\rightarrow$ \nutau \ oscillations. We have taken it that
the maximum energy attained by a SNR $\nu$ 
is 1 PeV ($\sim \ 1/12 \times 10^{16}$ eV).

Note that, above background and below the maximum neutrino energy, there are 
only two decade energy bins operating in a South Polar detector
for all three neutrino sources. IceCube might provide evidence for sub-dominant
oscillations only via diagnostic (2) -- observation of energy dependence
in the value of ${\Phi_{\nu_{\mu}}^{obs}}/{\Phi_{\nu_{\mu}}^{theor}}$.
We repeat, however, that positive identification of such variation
requires that $\Phi_{\nu_{\mu}}^{theor}$ be well determined which, in turn,
necessitates accurate determinations of the normalizing differential
photon flux at 10 GeV and the photon spectral index.

On the other hand, Sgr A East shows up in four energy bins for a Mediterranean
detector and the other two sources in three. Thus the possibility that 
the third oscillation diagnostic might be brought into play is held out. 
Whether this diagnostic can be made to work in practice depends on 
statistical error, granted that
 potential confounds like energy dependence in the
detector response are well-enough pinned down.

Figure (2) illustrates the expected, 10-year
event rates in a km$^3$ detector at
the Mediterranean for the most promising oscillation scenario --
six mass eigenstate mixing -- and the best
astrophysical neutrino source, Sgr A East. The figure details event rates in 
decade energy bins with no oscillations and also (atmospheric and) 
sub-dominant 
oscillations governed by \dmsub's from $10^{-14}$ to $10^{-9} \ eV^2$.
The error bars indicate the statistical error ($\sqrt n/n$) in each `datum'.


\begin{figure}[thb]\label{fig:events}   
 {\begin{turn}{270}
\centerline{\plotone{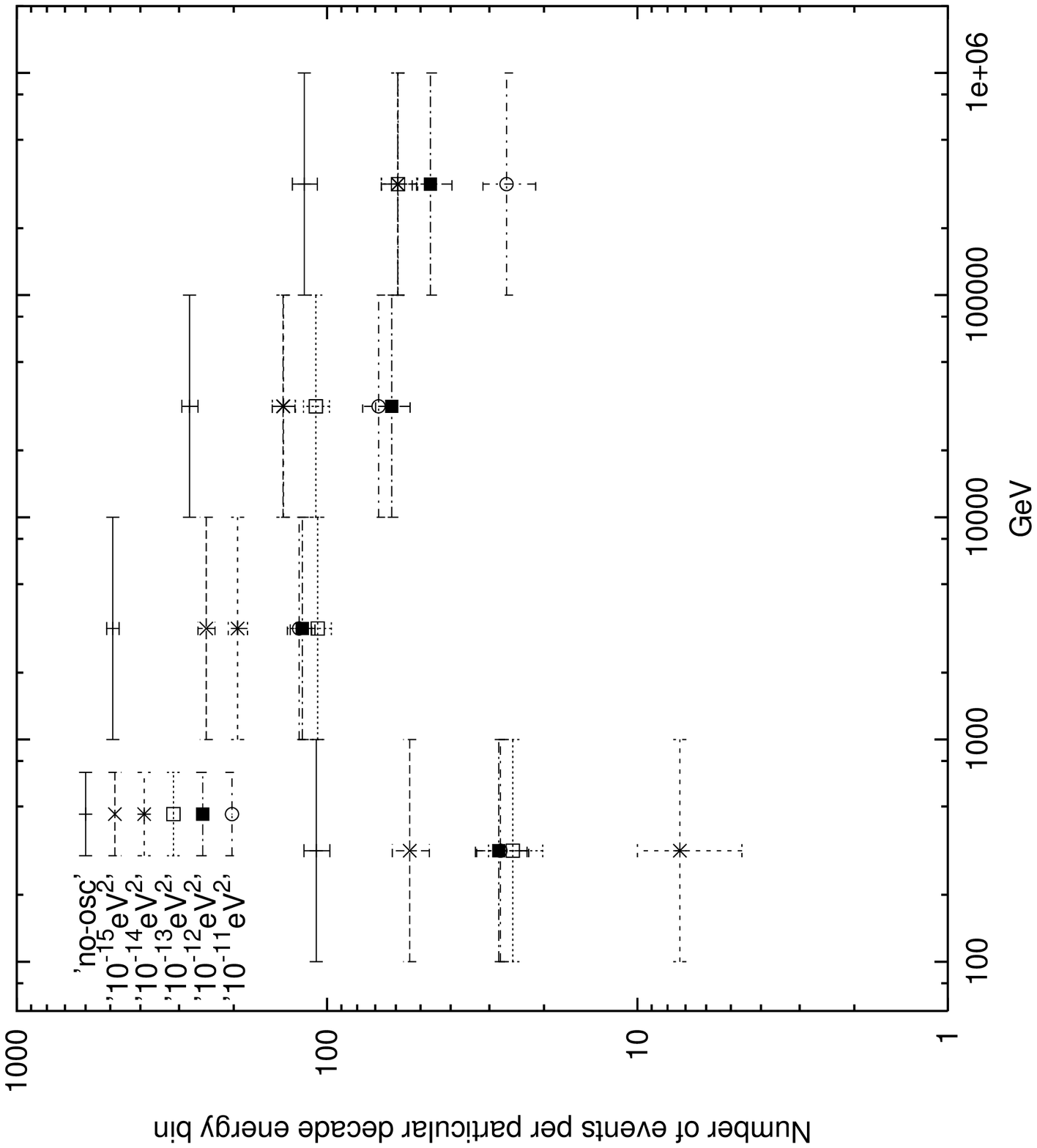}}
 \end{turn}}
\caption{Event rates in decade energy bins due to the Sgr A East $\nu$ source
with a number of different oscillation scenarios. The horiontal 
error bars indicate
the range of energy which defines each of the four active bins.
}
\end{figure}

\noindent From Figure (2) one may conclude that \dmsub's larger than 
$\sim 10^{-11} \ eV^2$ are not discernible one from another. At \dmsub \
values of $10^{-9}$ oscillation are completely averaged over Sgr A East's 
observable neutrino spectrum.

Figure (3) illustrates a way to test the hypothesis that a neutrino spectrum
obeys a power law. The $E^2$-weighted 
differential flux at each particular energy $E$
is related to the integral flux from $E$ to $10E$, $N(E,10E)$, 
one would infer from
the event rate (in that particular decade energy bin) by the relation
\begin{equation}
\frac{dN(E)}{dE} \ = \ \frac{1 - \alpha}{10^{1-\alpha} - 1} 
\frac{N(10E,E)}{E}.
\end{equation} 
The figure shows 
the differential flux at the minimum energy
of each of the four decade energy bins one might extrapolate from 
actual observation over ten years of Sgr A East with a km$^3$ neutrino
telescope at Mediterranean latitude, assuming a power-law spectrum.
The error bars are calculated from the relative statistical
error in the event rate for the same energy bin. 
{\it If} the plotted points form a straight line within statistical error, 
{\it then} the power law hypothesis is consistent with the `data'. 
From Figure (3), however, 
one may see by eye immediately that for a range of \dmsub \ values, a
constant power-law (i.e., straight line)
fit to the differential flux is {\it not} possible within
the error bars.


\begin{figure}[thb]\label{fig:sgrdifflux}
 {\begin{turn}{270}
\centerline{\plotone{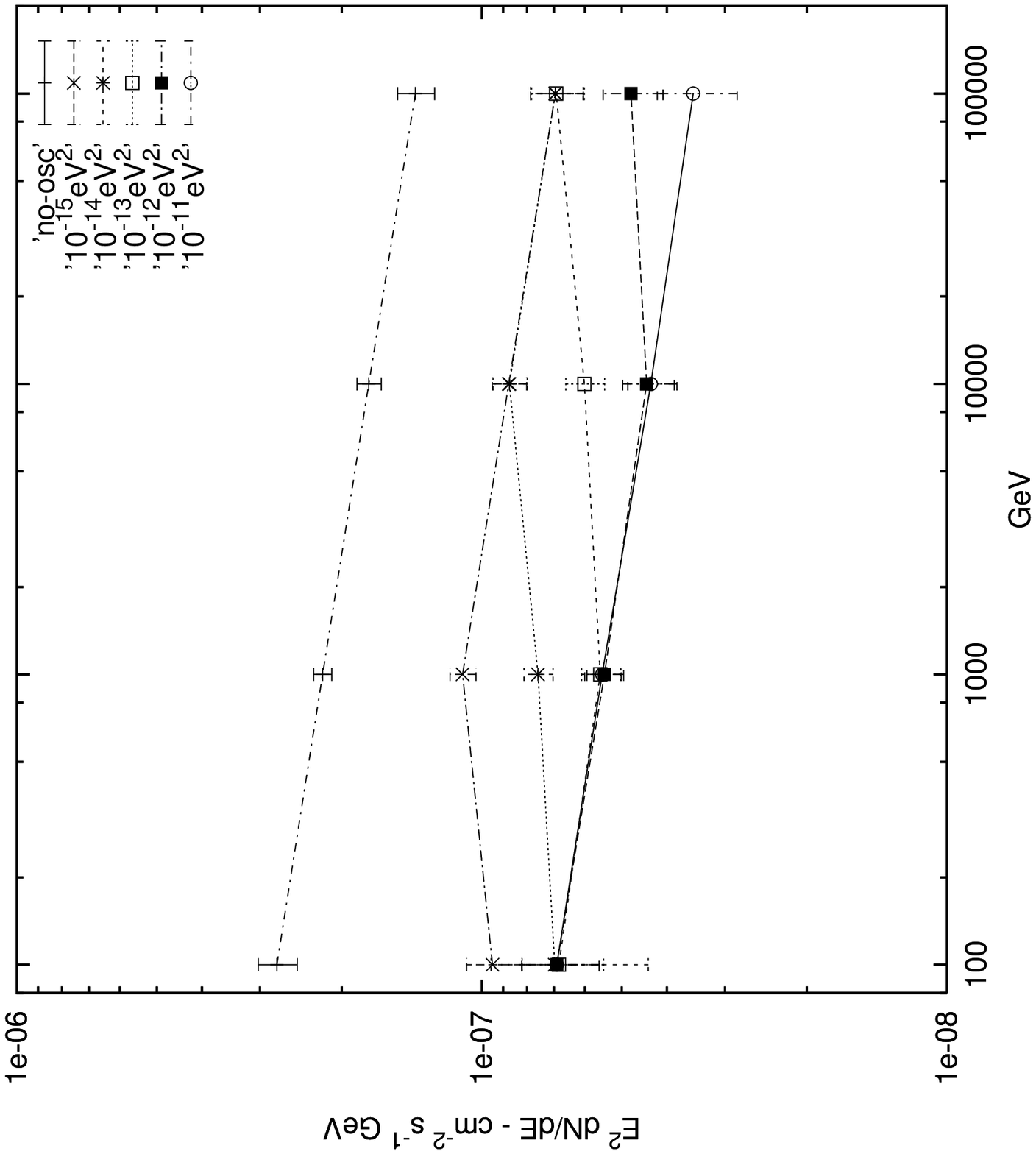}}
 \end{turn}}
\caption{Energy$^2$-weighted differential flux from the
Sgr A East $\nu$ source
with a number of different oscillation scenarios. See the text for a 
full description of the method used to generate this figure.
}
\end{figure}

Employing a least-squares fit to the differential flux data for the
various values of \dmsub \ with a two-free-parameter function,
$f(E) = a \times E^{-b}$ we determine that in every case the fitting algorithm
settles to a value for $b$ {\it less} than 2. Such a flat spectrum is
never observed in nature nor is it predicted by shock acceleration models. 
In fact, a power law going with index -2 describes the flattest reasonable
spectrum. 
We therefore fit to a one-free-parameter function, $f(E) = a \times E^{-2.0}$.
The $\chi^2$ values determined by such a fit are presented in Table (2).

\begin{table}[h]
\begin{tabular}{|cc|} \hline
$\delta m_{SUB}$ (eV)& $\frac{\chi^2}{d.o.f.}$  \\ 
\hline
$10^{-11} $ & $\sim$ 0\\
$10^{-12} $ & 3.2 \\
$10^{-13} $ & 16.1 \\
$10^{-14} $ & 19.3 \\
$10^{-15} $ & 12.0\\
 & \\
no osc  & $\sim$ 0\\
\hline
\end{tabular}
\caption{$\frac{\chi^2}{d.o.f.}$ values for the
fitting of a power law of index 2.0 to modelled differential
flux determinations for different values of \dmsub.
}
\end{table}

\vspace{0.5cm}
Note that this oscillation diagnostic works fairly independently of the
overall normalization of the flux, i.e., one simply needs {\it enough} events
from the neutrino
source
without having to know exactly its flux in the absence of oscillations. 
As the table and figure above illustrate,
Sgr A East does, indeed, produce sufficient events over ten years
for this method. This cannot, unfortunately, be said for the other two SNRs 
under consideration. Figure (4) shows the differential neutrino
flux from SN 1006 for the three active angle bins in a Mediterranean
km$^3$ detector (generated in the same fashion as for Figure(3) above). 

\begin{figure}[thb]\label{fig:1006difflux}
 {\begin{turn}{270}
\centerline{\plotone{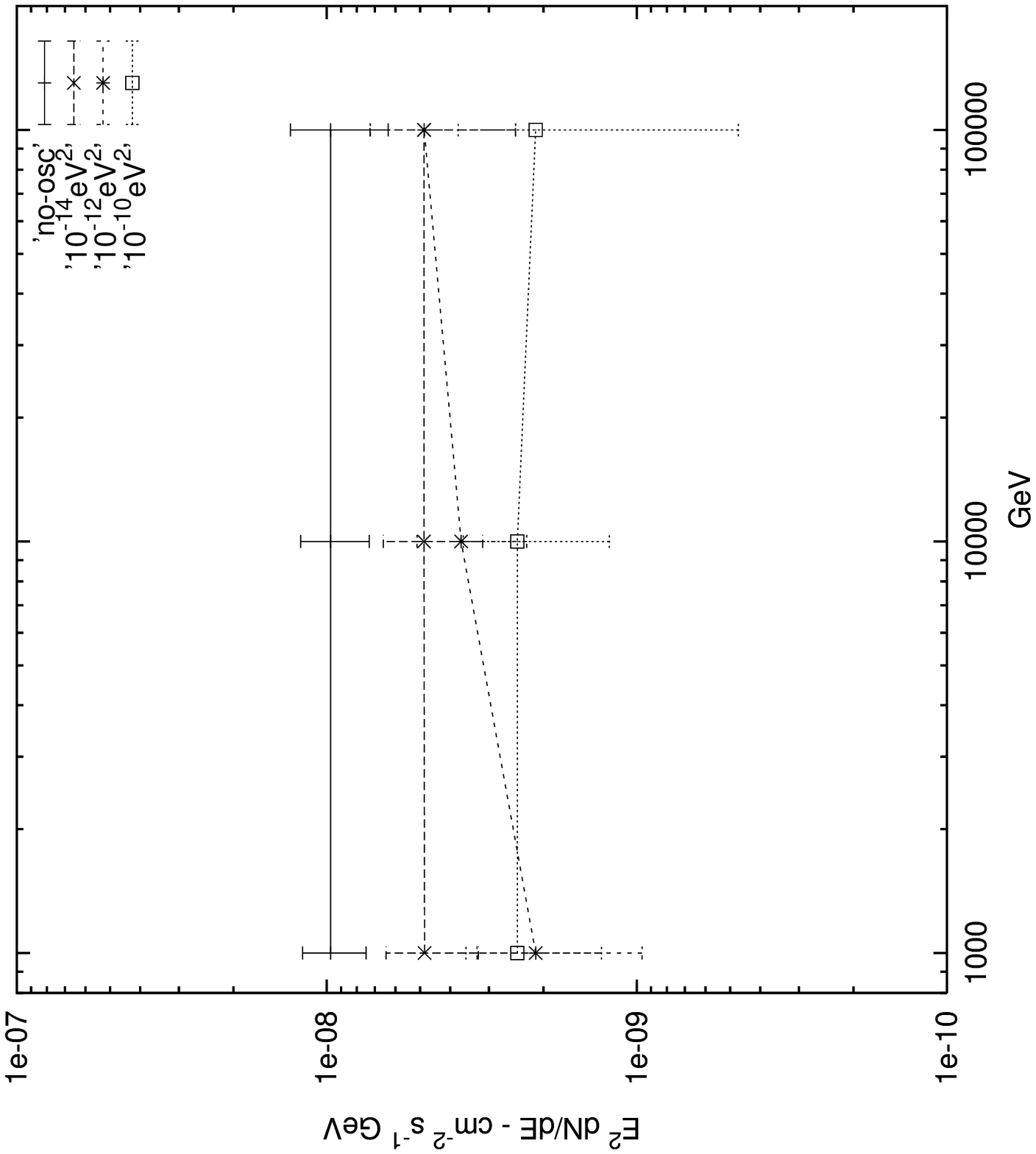}}
 \end{turn}}
\caption{Energy$^2$-weighted differential flux from the
 1006 $\nu$ source
with a number of different oscillation scenarios. The figure
is generated in the same fashion as Figure (3); see the text for details.}
\end{figure}
\noindent Again by eye, one may note from Figure (4)
that the unavoidable statistical
error (calculated for ten years' events) washes out the oscillation 
signature. In fact, the no oscillation case may only just 
be distinguished from all the separate oscillation examples.
The neutrino signal from 
SN 1006 (and also that from 
SNR RXJ1713.7-3946 and similar, yet undiscovered objects), therefore, 
requires either or both longer observation periods than a decade or 
larger than 1 km$^3$ detectors to be useful in providing 
evidence for spectral distortion due to sub-dominant oscillations. 
It may be that the signals from a number of such similar SNRs are able to
be statistically combined to provide a data set large enough to be probed for 
sub-dominant oscillations. Further, observation 
(or, indeed, non-observation) of neutrino signals from
these objects will settle once and for all the question of whether it
is hadronic or leptonic acceleration that is ultimately responsible
for their high energy $\gamma$-ray signals and will, therefore, be of crucial
import to cosmic ray research.

\section{Conclusion}

We have described a general technique via which the effects
of sub-dominant neutrino
oscillations might be uncovered in the \numu \ spectra of galactic
SNRs. This technique does not require either observation of \nue's or
\nutau's or unrealistically constrained measurements of
these objects' high energy $\gamma$-ray signals. 
We have determined through careful modeling that, when applied to the 
Sgr A East neutrino signal, the technique
will allow for the 
discovery or exclusion of sub-dominant neutrino oscillations governed by
a \dm \ parameter with a value in the range $10^{-12}$
to $10^{-15}$ eV$^2$. Such values cannot be probed by any conceivable
terrestrial and solar system neutrino experiments.

\section{Acknowledgments}

R.M.C. would like to thank R. Foot, 
F. Halzen, P. Lipari, S. Liu, D. Noone, A. Oshlack, and R. Protheroe
for useful discussion and 
correspondence. 
R.M.C. gratefully acknowledges the hospitality
shown him by F.M. and the University of Arizona; he is
supported by the Commonwealth of Australia.
R.R.V. is supported by the Australian Research Council.  
F.M. is partially supported by NASA
under grant NAGW-2518 at The University of Arizona.


\end{document}